\documentclass[conference]{IEEEtran}
\IEEEoverridecommandlockouts
\usepackage{cite}
\usepackage{amsmath,amssymb,amsfonts}
\usepackage{algorithmic}
\usepackage{graphicx}
\usepackage{textcomp}
\usepackage{xcolor}
\def\BibTeX{{\rm B\kern-.05em{\sc i\kern-.025em b}\kern-.08em
    T\kern-.1667em\lower.7ex\hbox{E}\kern-.125emX}}
\begin{document}

\title{Automated Personnel Selection for Software Engineers Using LLM-Based Profile Evaluation\\
}

\author{
\IEEEauthorblockN{Ahmed Akib Jawad Karim}
\IEEEauthorblockA{\textit{Dept. of CSE} \\
\textit{BRAC University}\\
Dhaka-1212, Bangladesh \\
ext.akib.jawad@bracu.ac.bd}

\and
\IEEEauthorblockN{Shahria Hoque}
\IEEEauthorblockA{\textit{Dept. of CSE} \\
\textit{BRAC University}\\
Dhaka-1212, Bangladesh \\
shahriahaquehridoy@gmail.com}
\and
\IEEEauthorblockN{Md. Golam Rabiul Alam}
\IEEEauthorblockA{\textit{Dept. of CSE} \\
\textit{BRAC University}\\
Dhaka-1212, Bangladesh \\
rabiul.alam@bracu.ac.bd}
\and
\IEEEauthorblockN{Md. Zia Uddin}
\IEEEauthorblockA{\textit{Software and Service Innovation Group} \\
\textit{SINTEF Digital, Oslo, Norway} \\
zia.uddin@sintef.no}
}

\maketitle

\begin{abstract}
Organizational success in today’s competitive employment market depends on choosing the right staff. This work evaluates software engineer profiles using an automated staff selection method based on advanced natural language processing (NLP) techniques. A fresh dataset was generated by collecting LinkedIn profiles with important attributes like education, experience, skills, and self-introduction. Expert feedback helped transformer models—including RoBERTa, DistilBERT, and a customized BERT variation, LastBERT—to be adjusted. The models were meant to forecast if a candidate's profile fit the selection criteria, therefore allowing automated ranking and assessment. With 85\% accuracy and an F1 score of 0.85, RoBERTa performed the best; DistilBERT provided comparable results at less computing expense. Though light, LastBERT proved to be less effective, with 75\% accuracy. The reusable models provide a scalable answer for further categorization challenges. This work presents a fresh dataset and technique as well as shows how transformer models could improve recruiting procedures. Expanding the dataset, enhancing model interpretability, and implementing the system in actual environments will be part of future activities.

\end{abstract}


\section{Introduction}

Selecting the right personnel is a critical determinant of success for organizations across all sectors. With technological advancements, increased competition, and a rising pool of skilled candidates, the importance of a systematic and efficient recruitment and selection process has become paramount \cite{makhijani2020automated}. Organizations today face the dual challenge of not only attracting but also retaining highly qualified employees who align with their strategic objectives. This is particularly important in a competitive job market where the placement of skilled workers can significantly impact a company's ability to maintain high performance and achieve sustainable growth.

An efficient recruitment and selection (R\&S) approach guarantees that companies can find staff members with the required competencies to fulfill corporate objectives at the lowest feasible cost \cite{hakkani2019nlp}. Particularly in information-driven industries like software engineering, the need for experts with particular abilities like programming knowledge, business acumen, and domain-specific knowledge is fast expanding. To simplify the choosing process and make data-driven choices, companies are turning to automated technologies and artificial intelligence-driven models more and more.

Emphasizing key components such as technical skills, educational background, job experience, and self-introductions, we collect and assess LinkedIn profiles of software engineers to provide a new dataset for this research. Adding expert evaluations supplied by business experts, which serve as a benchmark for candidate assessment, helps us to enhance the dataset even further. Our approach automatically predictions scores based on these criteria using state-of- the-art transformer models: RoBERTa, DistilBERT, and a besised BERT version. Offering a scalable people selection plan, the trained models are kept and might be used once again for next projects.


\textbf{The main contributions of this study are:}
\begin{itemize}
    \item Produced a unique dataset with profiles of software engineers combining evaluations from professionals with a broad spectrum of traits.
    \item  To improve candidate ranking and selection and therefore enhance the whole decision-making process, integrated the TOPSIS (Technique for Order of Preference by Similarity to Ideal Solution) approach.
    \item Trained and fine-tuned advanced language models including RoBERTa, DistilBERT, and LastBERT to automatically classify profiles as suitable or unsuitable. The models were saved and deployed for future use, providing a reusable foundation for automated personnel selection.

\end{itemize}

\section{Related Work}

The field of automated personnel selection has seen significant advancements, particularly with the integration of natural language processing (NLP) techniques and machine learning models. Nevertheless, there is little research on the use of advanced language models like RoBERTa and DistilBERT particularly for the recruitment of software developers. Several studies have used TOPSIS and the Analytical Hierarchy Process (AHP) among conventional decision-making tools in personnel selection. To evaluate candidates for a smart village project in Cairo, Egypt, Liu et al. \cite{liu2020integration} for example coupled AHP and TOPSIS. Although this method was successful for multi-criteria decision-making, its scalability and automation for big-scale people selection were restricted as it mostly depended on hand expert input. Likewise, Hsu et al. \cite{hsu2019fuzzy} enhanced candidate assessments using the FANP (Fuzzy Analytical Network Process) technique so strengthening the dependability of the decision-making process. These conventional methods still lack, meantime, the automation and adaptability provided by contemporary NLP methods and transformer models. FANP provided more consistent performance than conventional AHP models, but the process was still manual and depended on expert opinions missing automation and scalability. Transformer-based models have made great advancements in sentiment analysis and natural language processing. Research such as that by Sun et al. \cite{sun2019fine} showed how well fine-tuning BERT variants for sentiment analysis tasks performed on many benchmarks. Liu et al. \cite{liu2019roberta} first presented RoBERTa, which enhanced BERT by eliminating the Next Sentence Prediction (NSP) goal and using dynamic masking, therefore producing better outcomes in several text classification tasks with accuracies between 86\% and 90\%. Despite these advancements, the application of such models in personnel selection remains limited. Most existing works rely on basic feature extraction from resumes or LinkedIn profiles \cite{raman2022linkedin}. For example, Kausar et al. \cite{kausar2021role} explored the use of LinkedIn data for candidate selection but did not incorporate advanced deep learning models for automation. Additionally, traditional studies like those by Zuo et al. \cite{zuo2020evaluation} used fuzzy logic and MATLAB-based models for personnel evaluation in specialized sectors, but these methods lack the generalizability and adaptability of transformer-based models.

While NLP techniques have made tremendous progress in automated staff selection, little research has been done on the use of advanced transformer models for selecting software engineers. Traditional systems abound in conventional decision-making models and manual methods, therefore limiting scalability and automation. This paper addresses these constraints by using a totally automated staff selection process tailored for software engineers, a distinctive LinkedIn profile dataset, expert evaluations, and transformer models like as RoBERTa and DistilBERT. The sections that follow demonstrate the results of our approach and go into our methodology.

\section{Methodology}

\subsection{Dataset}
This study's dataset was created from scratch by collecting profile data from 100 LinkedIn profiles of software engineers. The dataset focused on four key categories: experience (years), education, skills, and a self-written "About" section. Experience data included the number of years engineers worked at various companies. Educational details captured graduation years and achievements, while skills data focused on programming and technical expertise. The "About" section provided insights into self-perception and personal interests. Additionally, an "Overall" score was included based on evaluations by senior professionals to help select candidates for senior software engineer roles. Given the inherent imbalance in the dataset, preprocessing and augmentation techniques were applied. The data was split into training (80\%) and testing (20\%) sets. 

Candidate selection was performed using the TOPSIS method, which involved calculating a normalized decision matrix, followed by a weighted matrix, and determining the ideal best and worst values. We computed the Euclidean distances from these values to derive the performance scores. The model's performance was further validated using metrics like RMSE, MAE, Manhattan Distance, and Cosine Similarity. 

\subsection{Data Preprocessing}
The initial preprocessing involved several steps:
\begin{itemize}
    \item \textbf{Synonym Replacement:} To enhance data diversity, a synonym replacement technique was applied, where words in the original text were replaced with their synonyms using WordNet. This step introduced variability in the training set, helping the model generalize better.
    \item \textbf{Sentiment Label Mapping:} The overall sentiment scores were mapped to binary labels. Scores ranging from 0 to 2 were labeled as negative (0), while scores of 3 and above were labeled as positive (1).
    \item \textbf{Data Augmentation:} The synonym-augmented text was combined with the original dataset, effectively increasing the number of training samples.
    \item \textbf{Class Balancing:} The dataset was imbalanced, with one class having more samples than the other. We addressed this issue by resampling the minority class to match the majority class size, ensuring a balanced dataset.
\end{itemize}

After preprocessing, the augmented dataset was split into training and testing sets with an 80-20 split, ensuring sufficient data for model evaluation.
\subsection{Feature Analysis}
\begin{figure}[htbp]
    \centering
    \includegraphics[width=0.4\textwidth]{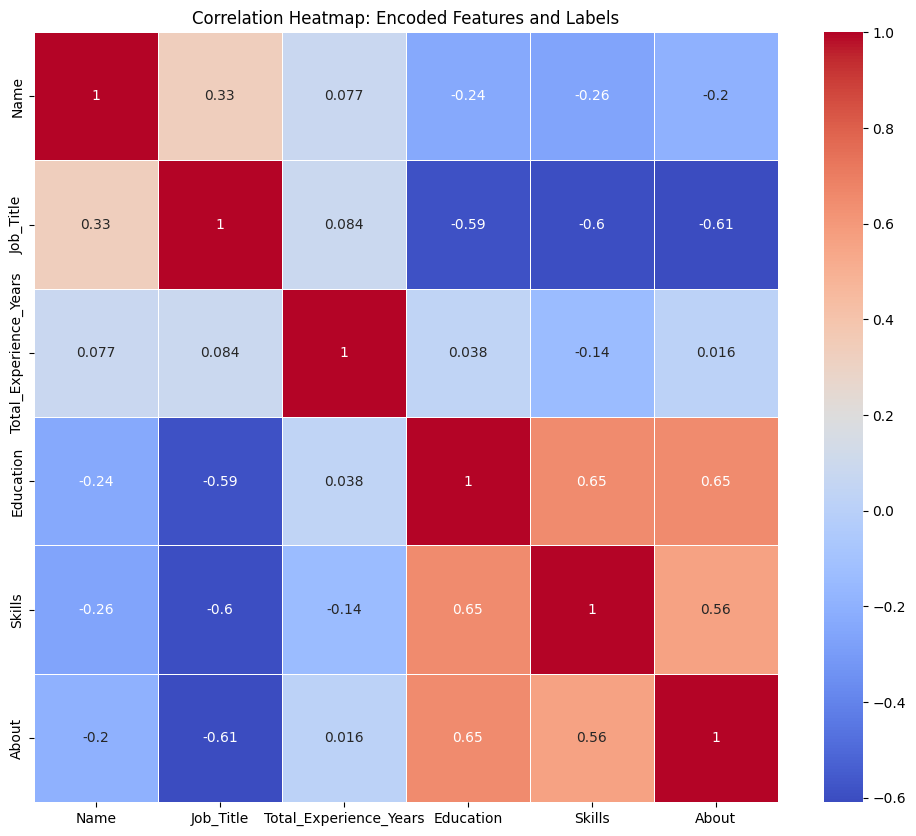}
    \caption{Correlation Heatmap: Encoded Features and Labels}
    \label{fig:correlation_heatmap}
\end{figure}

A correlation analysis was conducted to examine relationships between features before model training. The correlation heatmap in Figure~\ref{fig:correlation_heatmap} highlights key patterns. `Education` and `Skills` show a strong positive correlation (0.65), indicating that better-educated individuals often have broader skill sets. The `About` section is also positively correlated with both `Education` (0.65) and `Skills` (0.56), linking well-written profiles with higher qualifications. Conversely, `Job\_Title` negatively correlates with both `Education` (-0.59) and `Skills` (-0.60), suggesting that higher-ranking titles may require fewer qualifications. These insights guided feature prioritization.

\subsection{Training Methodology}
Three pre-trained transformer models—RoBERTa, DistilBERT, and LastBERT—were fine-tuned for sentiment classification on the prepared dataset. Each model was trained separately using the Hugging Face \texttt{Trainer} API with the following setup:

\subsubsection{RoBERTa}
The \texttt{roberta-base} model and tokenizer were used for sequence classification. Text sequences were tokenized using padding and truncation to a fixed length. The training was performed with the following hyperparameters: 15 epochs, a batch size of 16, a learning rate of 2e-5, 1000 warmup steps, a weight decay of 0.01, and early stopping with patience of 3 epochs.

\subsubsection{DistilBERT}
The sequence classification challenge used the \texttt{distilbert-base-uncased} model and tokenizer. In order to tackle the problem of class imbalance, class weights were included in the procedure. The class weights were derived by evaluating the distribution of sentiment classes in the training dataset and were used throughout the model training process. The training configuration included 15 epochs, a batch size of 16, a learning rate of 2e-5, 1000 warmup steps, a weight decay of 0.01, and early stopping with the patience of 3 epochs. Following the completion of the training phase, a revised classification threshold was used to increase the sensitivity towards the negative class.

\subsubsection{LastBERT}
The \texttt{LastBERT} model\cite{karim2024lastbert}, a variant of BERT, was optimized by lowering the number of layers and attention heads to decrease the computational load. The model was trained using class weights that are similar to those used in the DistilBERT model. The hyperparameters included 15 epochs, a batch size of 16, a learning rate of 2e-5, 1000 warmup steps, a weight decay of 0.01, and early stopping with a patience of 3 epochs. Additionally, the receiver operating characteristic (ROC) curve was analyzed after the training phase in order to determine the optimal classification threshold. Subsequently, this criterion was used to optimize the classification's effectiveness.

\subsection{Evaluation and model saving}
Every model was evaluated on the test dataset using measures including accuracy, F1-score, precision, and recall. A confusion matrix was created to visually demonstrate how effectively the model categorized positive and negative emotions. For LastBERT, a Receiver Operating Characteristic (ROC) curve was generated; the appropriate threshold was discovered by finding the point optimizing both sensitivity and specificity. The better models were maintained finally for use.

\subsection{Work Flow Diagram}
\begin{figure}[htbp]
    \centering
    \includegraphics[width=0.5\textwidth]{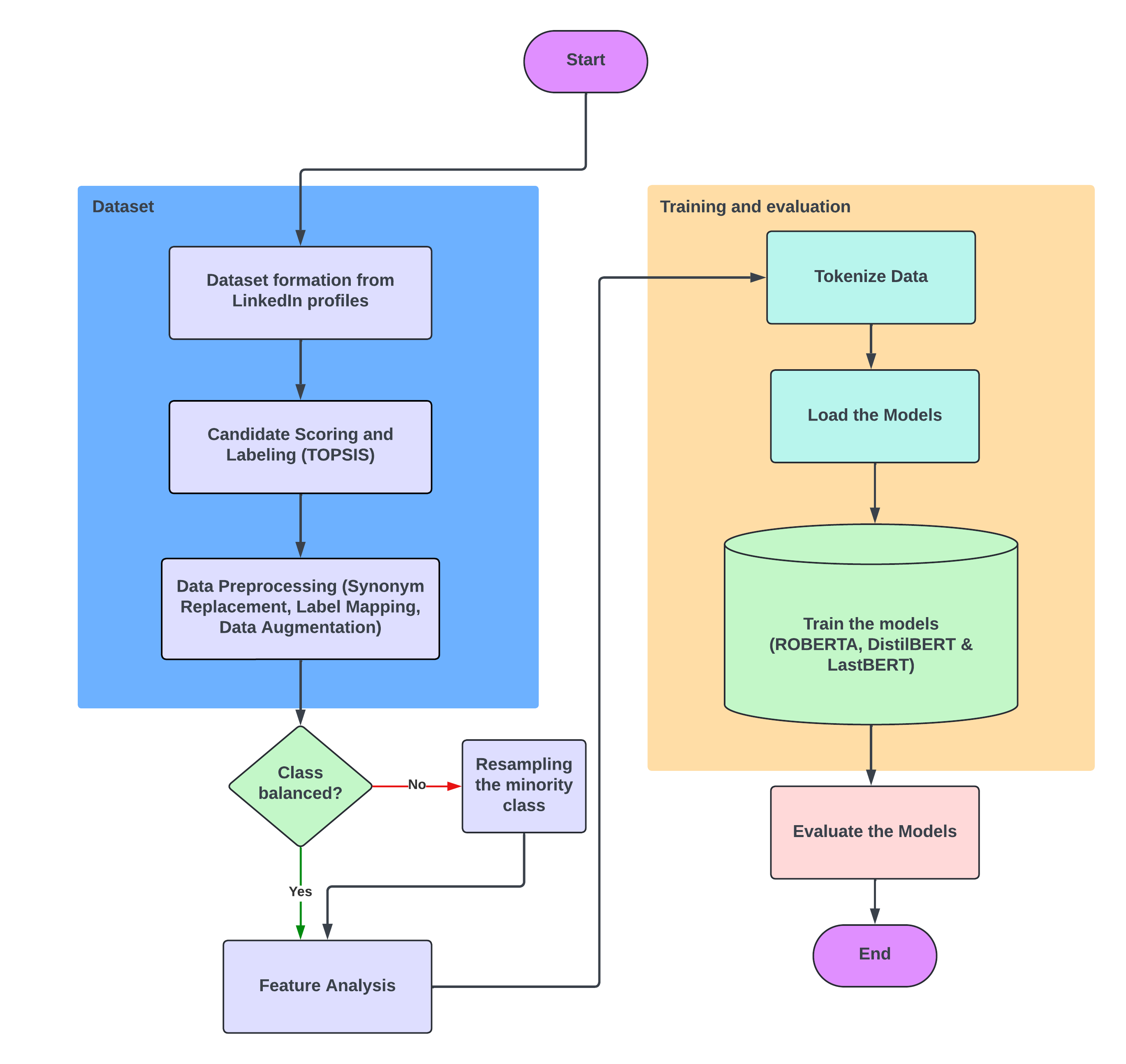}
    \caption{Top-level overview of the proposed personnel selection model}
    \label{fig:wf}
\end{figure}

The flow of this work—including model training, evaluation, and data preparation—is shown in 
Fig.~\ref{fig:wf}.

\section{Result Analysis}
We assess the DistilBERT, RoBERTa, and LastBERT models based on metrics like accuracy, F1 score, precision, recall, and optimal thresholds. Performance in sentiment detection is evaluated through training curves, confusion matrices, and ROC curves, offering a comparative overview of each model's suitability for this task while balancing sensitivity and specificity.

\subsection{TOPSIS Method}
The table~\ref{tab:topsis_results} summarizes key performance metrics from the TOPSIS method applied to our test dataset, which includes data from 20 individuals. The Root Mean Square Error (RMSE) is 7.503, indicating the selection model's error magnitude. The Mean Absolute Error (MAE) and Mean Absolute Percentage Error (MAPE) are 6.6 and 3.192\%, respectively, reflecting moderate deviations. The Manhattan Distance is 132, while the Cosine Similarity of 0.9843 shows high alignment between ideal and computed rankings. Normalized RMSE reduces to 0.1507, demonstrating the robustness of TOPSIS even after adjusting for scale differences across the features.

\begin{table}[htbp]
\centering
\caption{Performance Metrics for the TOPSIS Method}
\begin{tabular}{|c|c|}
\hline
\textbf{Metric} & \textbf{TOPSIS} \\ \hline
RMSE & 7.503 \\ \hline
Mean Absolute Error & 6.6 \\ \hline
Mean Absolute Percentage Error & 3.192 \\ \hline
Manhattan Distance & 132 \\ \hline
Cosine Similarity & 0.9843 \\ \hline
RMSE (Normalized Value) & 0.1507 \\ \hline
\end{tabular}
\label{tab:topsis_results}
\end{table}

\subsection{RoBERTa}

\begin{figure}[htbp]
    \centering
    \includegraphics[width=0.4\textwidth]{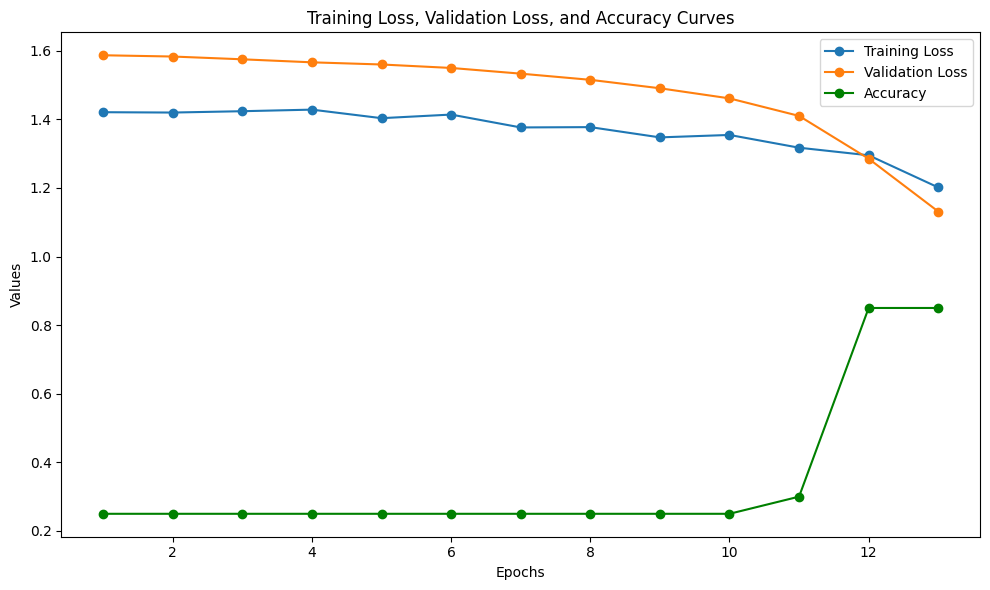}
    \caption{Training Loss, Validation Loss, and Accuracy Curves of RoBERTa}
    \label{fig:T_Roberta1}
\end{figure}

Besides the accuracy curve for the RoBERTa model, Figure~\ref{fig:T_Roberta1}, shows the training and validation loss curves. The training loss is constant at first, but from the ninth epoch, it starts to drop noticeably, suggesting better learning and parameter tuning. The validation loss shows a similar decreasing trend, suggesting the model is generalizing well to unseen data. Simultaneously, the accuracy curve exhibits a noticeable improvement after the 10th epoch, reaching approximately 85\%, indicating that the model’s performance stabilizes in the later epochs without signs of overfitting.

The RoBERTa model achieves a precision of 0.86, recall of 0.85, and F1 score of 0.85, effectively balancing false positives and false negatives. This consistent performance across metrics underscores the model's robustness for sentiment classification.

\begin{figure}[htbp]
    \centering
    \includegraphics[width=0.35\textwidth]{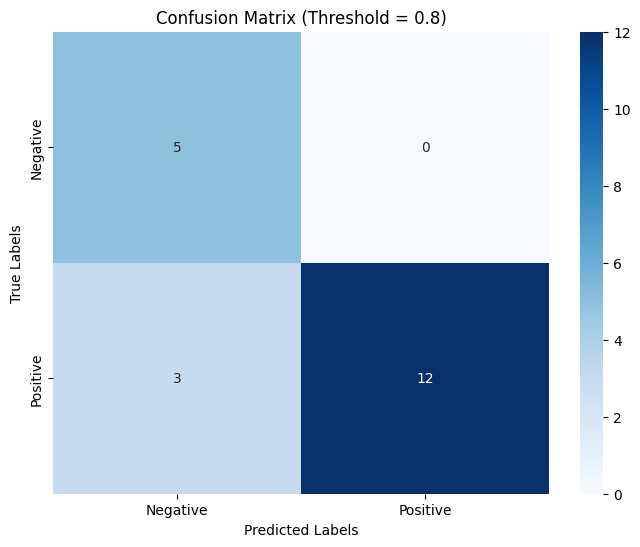}
    \caption{Confusion Matrix of RoBERTa (Threshold = 0.8)}
    \label{fig:CM_Roberta}
\end{figure}

The confusion matrix for RoBERTa, as shown in Figure~\ref{fig:CM_Roberta}, indicates that the model correctly predicted 5 negative and 12 positive samples. It misclassified 3 positive samples as negative while making no false positive predictions. This demonstrates the model's effectiveness in identifying positive cases, though a slight tendency exists to miss some, leading to false negatives. The overall accuracy remains high with minimal misclassifications.

\begin{figure}[htbp]
    \centering
    \includegraphics[width=0.3\textwidth]{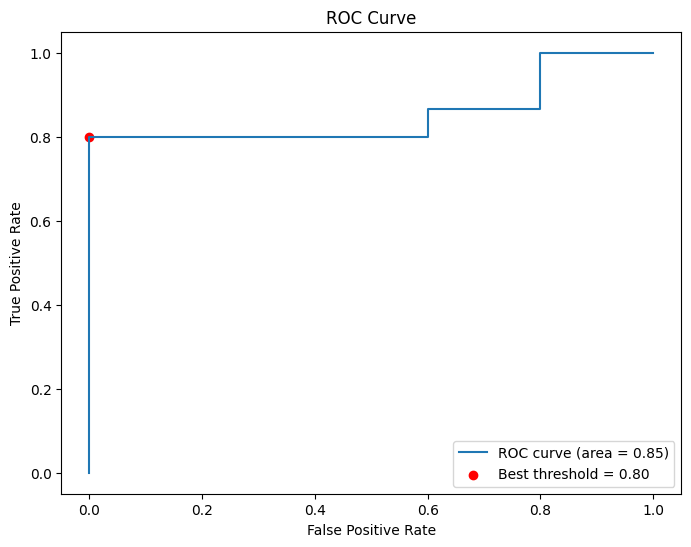}
    \caption{ROC Curve of RoBERTa (Best Threshold = 0.8)}
    \label{fig:ROC_Roberta}
\end{figure}

The ROC curve for the RoBERTa model, shown in Figure~\ref{fig:ROC_Roberta}, indicates an AUC of 0.85. The curve's steep rise shows effective distinction between positive and negative classes. The optimal threshold of 0.8, marked by the red point, balances sensitivity and specificity, validating the model’s performance.

\subsection{DistilBERT}

\begin{figure}[htbp]
    \centering
    \includegraphics[width=0.4\textwidth]{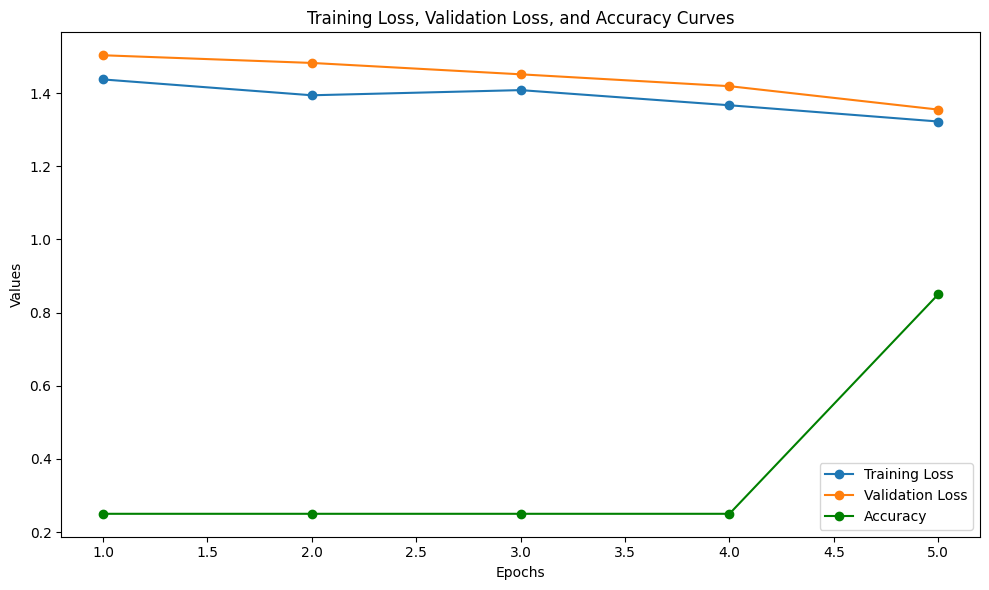}
    \caption{Training Loss, Validation Loss, and Accuracy Curves of DistilBERT}
    \label{fig:T_Distil}
\end{figure}

Figure~\ref{fig:T_Distil} shows the training and validation loss curves alongside the accuracy curve for the DistilBERT model. The training loss gradually declines, while the validation loss decreases more slowly, suggesting good generalization. A significant accuracy improvement is observed after the 4th epoch, reaching around 85\%, indicating the model needed several epochs to fine-tune and achieve optimal performance.

In terms of classification metrics, the DistilBERT model achieves a precision of 0.84, a recall of 0.85, and an F1 score of 0.84. These metrics indicate that the model performs consistently well, balancing false positives and false negatives effectively. The stable progression of these metrics underscores the model’s reliability for sentiment classification tasks.

\begin{figure}[htbp]
    \centering
    \includegraphics[width=0.35\textwidth]{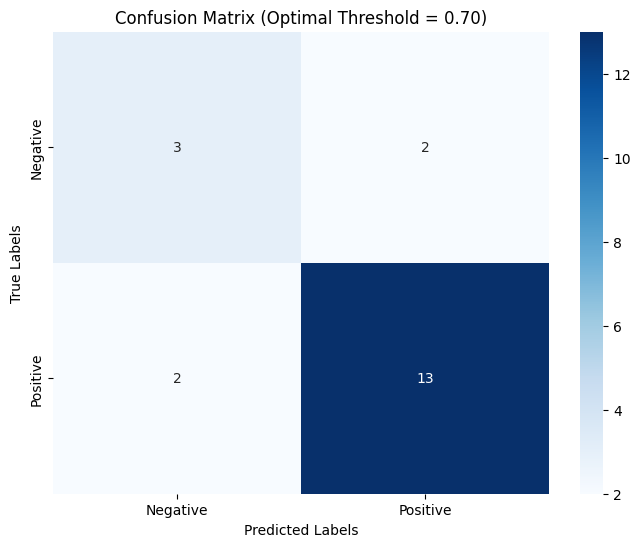}
    \caption{Confusion Matrix of DistilBERT (Optimal Threshold = 0.70)}
    \label{fig:CM_DistilBERT}
\end{figure}

The confusion matrix for DistilBERT, shown in Figure~\ref{fig:CM_DistilBERT}, demonstrates reasonable performance in distinguishing between positive and negative classes. It correctly classifies 13 positive and 3 negative samples, with 2 misclassifications for both classes. The model slightly favors the positive class, leading to a balanced distribution of false positives and false negatives.

\begin{figure}[htbp]
    \centering
    \includegraphics[width=0.3\textwidth]{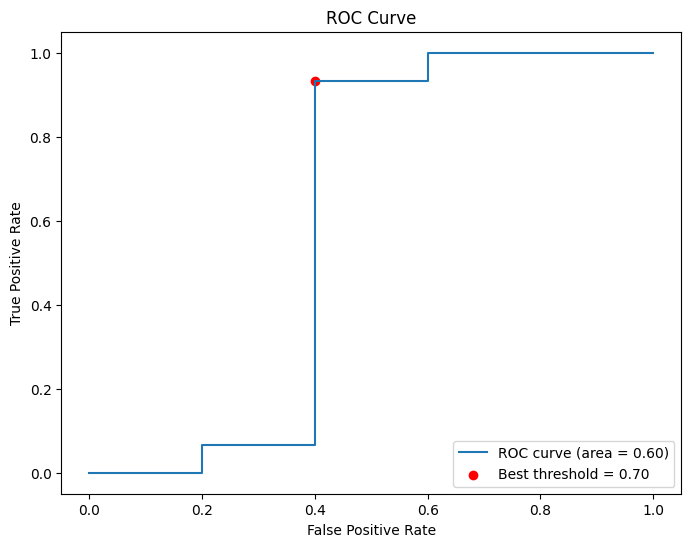}
    \caption{ROC Curve of DistilBERT (Best Threshold = 0.70)}
    \label{fig:ROC_DistilBERT}
\end{figure}

The ROC curve for DistilBERT in Figure~\ref{fig:ROC_DistilBERT} shows an AUC of 0.60, indicating moderate discriminative ability. The optimal threshold of 0.70, marked on the curve, balances sensitivity and specificity without favoring either class excessively.

\subsection{LastBERT}

\begin{figure}[htbp]
    \centering
    \includegraphics[width=0.35\textwidth]{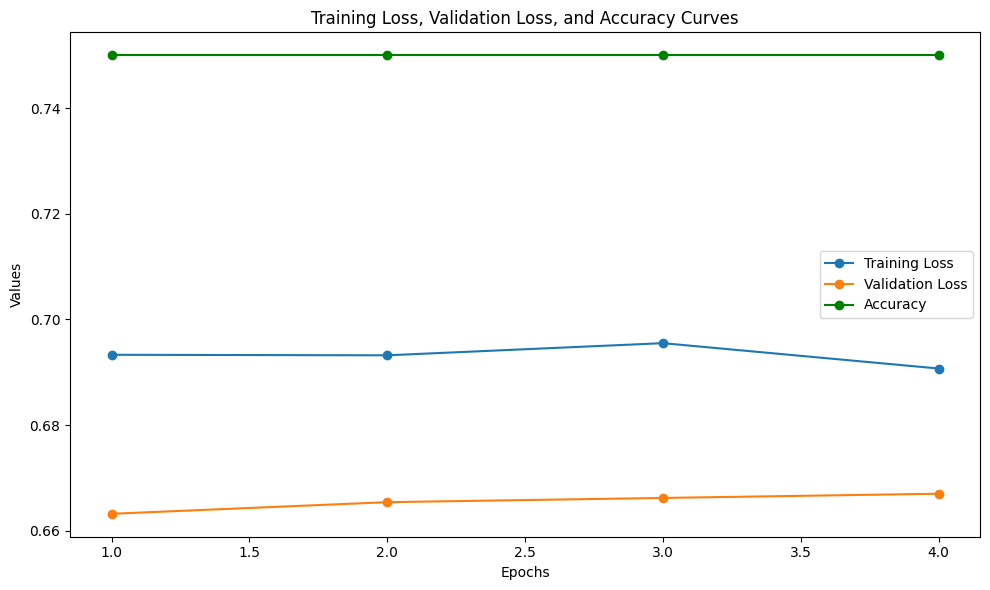}
    \caption{Training Loss, Validation Loss, and Accuracy Curves of LastBERT}
    \label{fig: T_LastBERT}
\end{figure}

Figure~\ref{fig: T_LastBERT} shows the training and validation loss curves alongside the accuracy curve for the LastBERT model. The training loss remains stable across epochs, indicating minimal overfitting or underfitting. However, the validation loss and accuracy show little change, with accuracy stabilizing around 75\%, suggesting limited learning capacity compared to RoBERTa and DistilBERT, both of which achieved higher performance in capturing dataset nuances.

The performance metrics for the LastBERT model indicate that precision remains around 0.82, recall at 0.75, and F1 score at 0.65, consistent with the results presented in Table~\ref{tab:model_comparison}. The stability of these metrics across epochs suggests that the model reaches its performance ceiling early in the training process, highlighting its limited capacity in feature extraction compared to RoBERTa and DistilBERT.

\begin{figure}[htbp]
    \centering
    \includegraphics[width=0.35\textwidth]{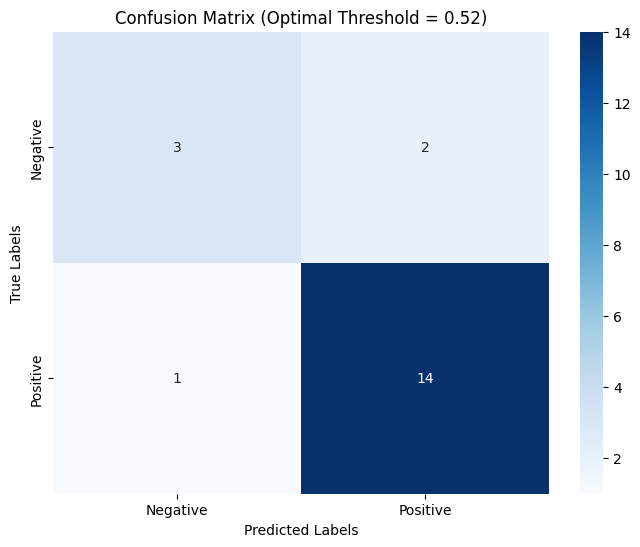}
    \caption{Confusion Matrix of LastBERT (Optimal Threshold = 0.52)}
    \label{fig:CM_LastBERT}
\end{figure}

The confusion matrix for LastBERT, shown in Figure~\ref{fig:CM_LastBERT}, shows a balanced performance. The model correctly classifies 14 positive and 3 negative samples, with only 1 positive sample misclassified as negative and 2 negative samples as positive. While slightly favoring the positive class, the model maintains a good balance between true positives and true negatives with minimal misclassifications.

\begin{figure}[htbp]
    \centering
    \includegraphics[width=0.3\textwidth]{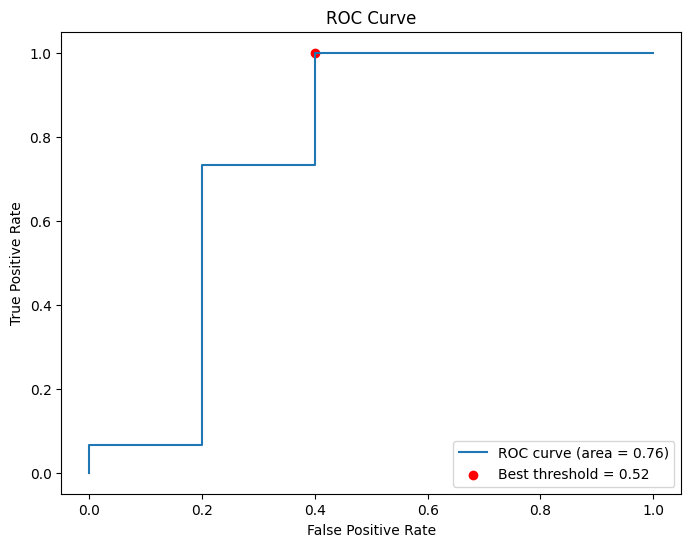}
    \caption{ROC Curve of LastBERT (Best Threshold = 0.52)}
    \label{fig:ROC_LastBERT}
\end{figure}

The ROC curve for LastBERT, shown in Figure~\ref{fig:ROC_LastBERT}, has an AUC of 0.76, indicating reasonable discriminative power but less robustness compared to other models. The optimal threshold of 0.52 balances sensitivity and specificity. While capable, LastBERT is less effective in precisely distinguishing between positive and negative sentiment for this dataset.

\subsection{Model Comparison}
\begin{table}[htbp]
\centering
\caption{Model Performance Comparison}
\begin{tabular}{|p{1.1cm}|p{1.7cm}|p{1.05cm}|p{0.9cm}|p{0.6cm}|p{1.25cm}|} %
\hline
\textbf{Model} & \textbf{\centering Accuracy (\%)} & \textbf{\centering F1 Score} & \textbf{\centering Precision} & \textbf{\centering Recall} & \textbf{Parameters} \\ \hline
RoBERTa & \centering 85.00 & \centering 0.85 & \centering 0.86 & \centering 0.85 & 125M \\ \hline
DistilBERT & \centering 85.00 & \centering 0.84 & \centering 0.84 & \centering 0.85 & 66M \\ \hline
LastBERT & \centering 75.00 & \centering 0.65 & \centering 0.82 & \centering 0.75 & 29M \\ \hline
\end{tabular}
\label{tab:model_comparison}
\end{table}

\begin{figure}[htbp]
    \centering
    \includegraphics[width=0.45\textwidth]{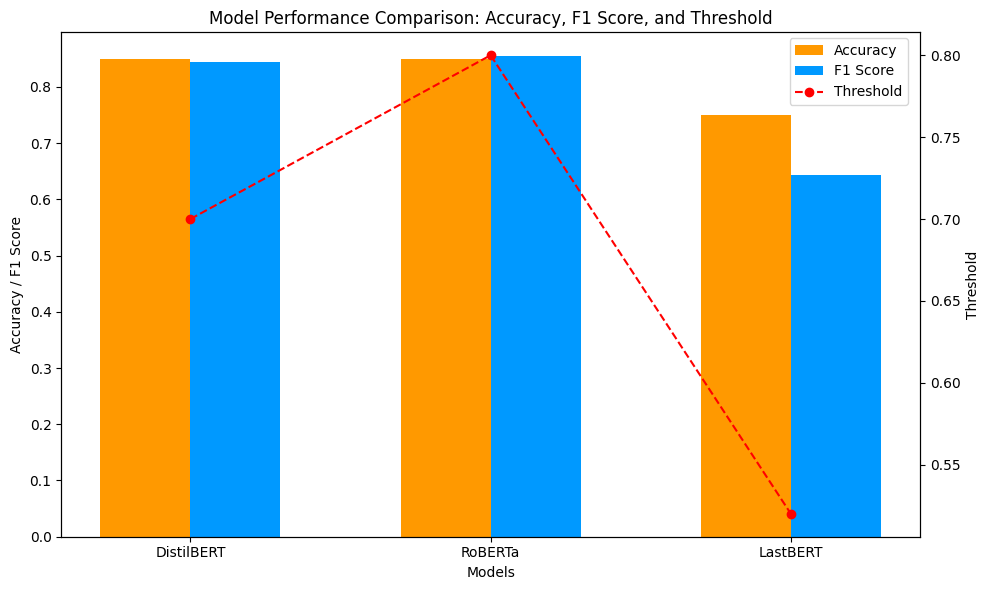}
    \caption{Model Performance Comparison: Accuracy, F1 Score, and Threshold}
    \label{fig:model_comparison}
\end{figure}

The comparison between DistilBERT, RoBERTa, and LastBERT is summarized in Table~\ref{tab:model_comparison} and visualized in Figure~\ref{fig:model_comparison}. Both DistilBERT and RoBERTa achieve similar accuracy (85\%), with RoBERTa slightly outperforming DistilBERT in F1 score (0.85 vs. 0.84) and precision (0.86 vs. 0.84). RoBERTa’s advantage can be attributed to its larger model size (125M parameters), allowing it to capture more complex data patterns. In contrast, LastBERT, with a smaller model size (29M parameters), shows a drop in performance, achieving 75\% accuracy and a 0.65 F1 score. It maintains a relatively high precision (0.82), reflecting a more conservative approach that reduces false positives but increases false negatives, leading to lower recall. The confusion matrices highlight these differences. RoBERTa shows only 3 false negatives and no false positives, confirming its strong performance. DistilBERT shows a balanced classification with 2 false positives and 2 false negatives. LastBERT misclassifies more samples with 2 false positives and 1 false negative, indicating its limited ability to capture nuanced patterns.

Figure~\ref{fig:model_comparison} emphasizes the trade-offs between accuracy, F1 score, and thresholds across the models. RoBERTa offers the best performance but with higher computational costs. DistilBERT provides a balanced trade-off with a smaller model size, while LastBERT, though lightweight, is less suitable for tasks requiring high accuracy.

\section{Conclusion and Future Work}

This paper presented an automated personnel selection system utilizing transformer-based models for profile analysis of software engineers' profiles. A novel dataset of LinkedIn profiles was created and enriched with expert evaluations for model benchmarking. RoBERTa, DistilBERT, and LastBERT were fine-tuned and evaluated across key metrics like accuracy, F1 score, precision, and recall. RoBERTa achieved the best performance with 85\% accuracy and a 0.85 F1 score, while DistilBERT offered comparable results with greater resource efficiency. LastBERT, though lightweight, demonstrated lower accuracy, making it better suited for scenarios with limited computational resources. The variation in optimal thresholds among models highlights how each model handles decision-making differently, underscoring the importance of customizing thresholds to balance sensitivity and specificity for specific applications. The findings confirm that transformer-based models can effectively automate personnel selection, offering a scalable and consistent alternative to manual processes. The saved models provide a reusable framework for future applications, potentially enhancing recruitment efficiency while minimizing bias.

Future work could involve expanding the dataset, integrating resume data for better generalizability, and exploring ensemble methods for improved accuracy. Additionally, deploying the models in real-world settings and integrating explainability techniques like SHAP or LIME would enhance model transparency and validate their effectiveness in practice.

In summary, this research lays the foundation for automated personnel selection using state-of-the-art NLP techniques, contributing to more data-driven and unbiased decision-making in human resource management.

\bibliographystyle{IEEEtran}
\bibliography{references}

\end{document}